\documentclass[sigconf]{acmart}

\usepackage{booktabs} 
\usepackage{graphicx}
\usepackage{subfigure}
\usepackage{epstopdf}

\usepackage{caption} 
\begin{document}

\title{User Donations in a Crowdsourced Video System}


\author{Adele Lu Jia}
\affiliation{%
  \institution{China Agricultural University}
  \city{Beijing}
  \country{China}
}
 \email{adele.lu.jia@gmail.com}
 
\author{Xiaoxue Shen}
\affiliation{%
  \institution{China Agricultural University}
  \city{Beijing}
  \country{China}
}
\email{xiaoxueshen@cau.edu.cn}

\author{Siqi Shen}
\affiliation{%
  \institution{National University of Defense Technology}
  \city{Changsha}
  \country{China}
 }
 \email{shensiqi@nudt.edu.cn}
 
 \author{Jun Xu}
\affiliation{%
  \institution{National University of Defense Technology}
  \city{Changsha}
  \country{China}
 }
 \email{junxu@nudt.edu.cn}


\begin{abstract}
Crowdsourced video systems like YouTube and Twitch.tv have been a major internet phenomenon and are nowadays entertaining over a billion users. 
In addition to video sharing and viewing, over the years they have developed new features to boost the community engagement and some managed to attract users to donate, to the community as well as to other users. 
User donation directly reflects and influences user engagement in the community, and has a great impact on the success of such systems. 
Nevertheless, user donations in crowdsourced video systems remain trade secrets for most companies and to date are still unexplored. 
In this work, we attempt to fill this gap, and we obtain and provide a publicly available dataset on user donations in one crowdsourced video system named \textit{BiliBili}. 
Based on information on nearly 40 thousand donators, we examine the dynamics of user donations and their social relationships, we quantitively reveal the factors that potentially impact user donation, and we adopt machine-learned classifiers and network representation learning models to timely and accurately predict the destinations of the majority and the individual donations.

\end{abstract}

\maketitle


\section{Introduction}

Crowdsourced video systems like YouTube and Twitch.tv nowadays entertain over a billion users and form a billion-dollar global industry. In addition to video sharing and viewing, many systems have introduced new features such as chat replay and social network incorporation to keep their proliferation. 
The companies behind, as well as some of the users, get in return huge revenues through advertisements and user donations. 
In either way, users are the main contributors except that in advertisements users act passively (by viewing the ads) whereas in donations users act proactively and voluntarily. While there are several analyses of advertisements in crowdsourced video systems \cite{VideoAdsIMC13, YouYubeAdsWebSci16}, user donation remains an unexplored area. 
In this article we conduct, to the best of our knowledge, the first in-depth analysis of user donations in crowdsourced video systems.

The main motivation behind is to fill the gap between the profound and the promising role that user donation plays in real crowdsourced video systems and the very limited understanding towards it in academia.
User donation in crowdsourced video systems is normally either made to the community, or to peer users through the contents (videos) they share.
Donations made to the community are often used for maintaining the community and for purchasing contents that the users desire.
Donations made to peer users, on the other hand, work as a reward and encourage them to contribute more contents.
User donation is one of the ultimate contributions that users can perform. It directly reflects and influences user engagement in the community, and therefore has a great impact on the success of the systems. 
In this work, we focus on the fundamental questions related to user donation and we seek to quantify the dynamics of user donations, to predict and to analyze the aspects that influence the destinations of the majority and the individual donations.

To this end, we have chosen BiliBili \cite{BiliBiliURL} as our research platform. BiliBili contains of a huge amount of User Generate Content (UGC) videos uploaded by its users and a small number of copyright video series purchased by the community administrators. Users in BiliBili can donate to other users, showing their support to others who have contributed UGC videos, as well as to the copyright video series, as compensation for the expense of the community administrators. 
Unlike YouTube and Twitch.tv wherein user donation remains a trade secret and the donation statistics are not publicly available, for each copyright video series and for each user, BiliBili displays the number of donations they have received, along with a list of the identities of the top 100 donators. By doing so, BiliBili creates a sense of friendly competition and encourages its users to donate.


Our analysis of BiliBili mainly consists of three parts:

\textbf{Measuring user donations.} In this work, we focus on user donations made to the copyright video series. Our dataset covers all the 2,678 copyright video series and 34,409 users who have donated to them. The information we obtained includes not only basic video characteristics like the duration and the popularity, but also user activities and interactions like who donates to which video series, how users follow and donate to each other, who uploads which UGC video and how these UGC videos perform. As our on-going work, we are collecting information on user to user donations (with over 10 million videos and over 200 million users). The dataset and analysis will be publicly available soon.

\textbf{Characterizing user donations.} We first quantitively reveal the scale and the characteristics of user donation in BiliBili by examining its copyright video series repository. We analyze the popularity and the number of donations the video series received, and we find them \textit{highly skewed}, with a small number of video series collecting a large portion of the total popularity and donation.
Then, we examine the follow, the upload and the donate activities of the users who have donated to the copyright video series. We find a long-tail distributed number of followers, a highly skewed number of donations made and received, and highly skewed popularities for the UGC videos they upload.
 
\textbf{Predicting user donations.} Applying our findings we answer two fundamental questions in user donation, i.e., \textit{where the majority of donations go} and \textit{who will donate to whom}. 
First, we build machine-learned classifiers to predict, based on the current information, which video series will attract most donations in the coming week. On a balanced dataset where random guessing would yield an accuracy of 50\%, our predictions achieve 97\%.
Secondly, we adopt heterogeneous network representation learning models to learn the embeddings of videos and users, based on which we predict and recommend, for each copyright video series, the potential users who will donate.

We summarize our contributions as follows:

\begin{itemize}

\item We collect, use, and for academic purposes offer public access\footnote{The dataset is available upon request via email to the first author.} to the dataset that contains detailed statistics for 2,678 copyright video series and 34,409 users who have donated to them (Section \ref{problem}).

\item We provide a characterization on user donation in BiliBili. Our analysis includes the statistical properties of copyright video series and the donation activity, the upload activity, and the follow activity of the users (Section \ref{char}).

\item We use network representation learning models and build machine-learned classifiers to identify with high accuracies the video series that will receive the majority donations, and to efficiently recommend, for each video series, potential users that will donate. (Section \ref{prediction}).

\end{itemize}

\begin{figure}
  \centering
    \includegraphics[width=0.4\textwidth]{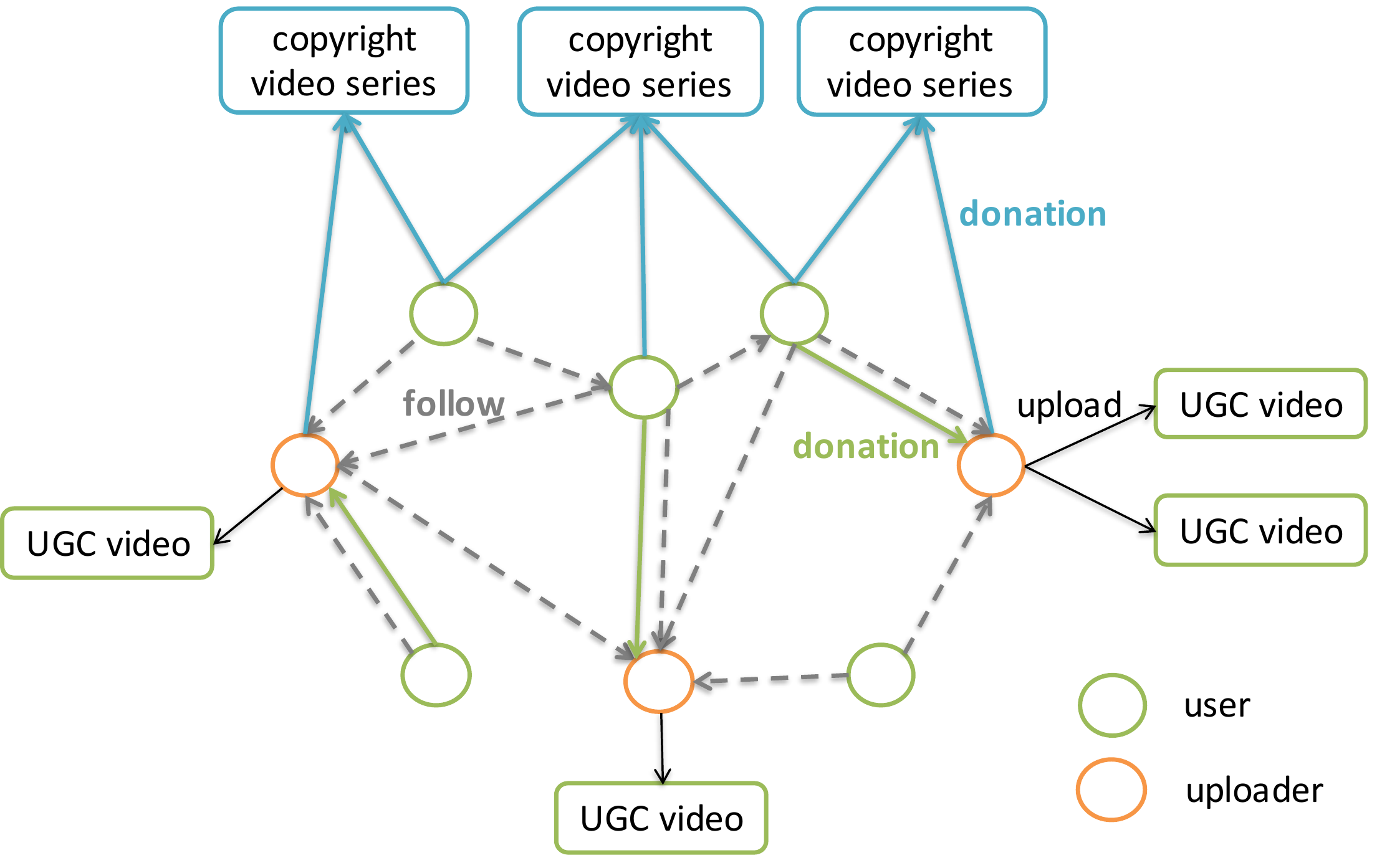}
  \caption{The BiliBili ecosystem}
  \label{BiliChengbao}
\end{figure}

\section{The BiliBili Donation Dataset}
\label{problem}

In this section, we give a brief introduction of the ecosystem of BiliBili, and we introduce our measurement methodology and the dataset used throughout this article.

\subsection{The BiliBili ecosystem}

BiliBili provides both video and social services. It mainly contains two types of videos, i.e., copyright video series purchased by the community administrators and UGC videos uploaded by the users. 
As in traditional video systems, users in BiliBili can view and share videos, vote and leave comments to videos, and subscribe to other users. BiliBili also provides several enhanced social features, such as user following and chat replay (named \textit{danmu} in BiliBili) wherein the chat from the past show up right next to and on top of the video for the current viewer, at exactly the video time when they were left. As a toy example, suppose that user A has left a comment when the video was on for 2 minutes. This comment will be displayed to all the other users who watch the video later when the video is on for 2 minutes for them. In this way, chat replay provides immersive viewing experiences and is adopted by a number of popular video systems including YouTube Live and Twitch.tv.

Being a crowdsourced video system, BiliBili managed to encourage their users to donate. Donations can be made in two ways, i.e., to the copyright video series and to individual users who have shared UGC videos. For each video series and for each user that receives donations, the number of donations they received, in total and for the past week, are highlighted in their homepages, along with a list of the top donators. 
Figure \ref{BiliChengbao} shows an overview of the BiliBili ecosystem.

\subsection{The BiliBili donation dataset}

\begin{table}
\small
\centering
\caption{Basic statistics of the BiliBili donation dataset}
\label{statistics_chengbao}
\begin{tabular}{l | l }
\hline
	$\#$ copyright video series & 2,786\\ \hline
 	$\#$ donators          & 34,409  \\     	
	$\#$ donators with UGC uploads & 6,876\\
	$\#$ UGC videos uploaded by the donators & 83,121\\
	$\#$ donators who received user donations & 1,085\\ 
	\hline
	$\#$ user-user follow      & 2,190,837\\
	$\#$ user-series donation (for the past week) & 44,916\\
	$\#$ user-user donation (for the past month) & 227\\
 \hline
\end{tabular}
\end{table}

BiliBili identifies each of its video and users with a unique number in the increasing order. Each identifier corresponds to a webpage with detailed video or user information that can be obtained with crawlers. 
In this work, we focus on donations made to copyright video series. We have obtained detailed information of all the 2,786 copyright video series (by the end of 2017), the 34,409 users who have donated to them at least once (named \textit{donators}), and the 83,121 UGC videos uploaded by these donators, which we name the \textit{copyright video series dataset}, the \textit{donator dataset}, and the \textit{UGC video dataset}, respectively.
As our on-going work, we are collecting statistics on user to user donations (with over 10 million videos and over 200 million users). The analysis and dataset will be publicly available soon.

\textit{The copyright video series dataset} contains, for each series, the number of views, the number of subscriptions, the number of danmus, the number of donations (in total and for the past week) it received, and the identities of the top 100 donators (can be cross-referenced with the donator dataset). 
\textit{The donator dataset} contains, for each donator, his followers, his followees, the repository of his uploaded videos (can be cross-referenced with the UGC video dataset), the number of the donations he \textit{received} (in total and for the past month), and the users who have donated to him for the past month.
\textit{The UGC video dataset} contains, for each video uploaded by the donators, the duration, the number of views, the number of favorites, the number of danmus, and the number of comments it collected. 

The basic statistics of our datasets are depicted in Table \ref{statistics_chengbao}.

\section{Characteristics of donations in BiliBili}
\label{char}

\begin{figure}
    \centering
    \includegraphics[width=0.4\textwidth]{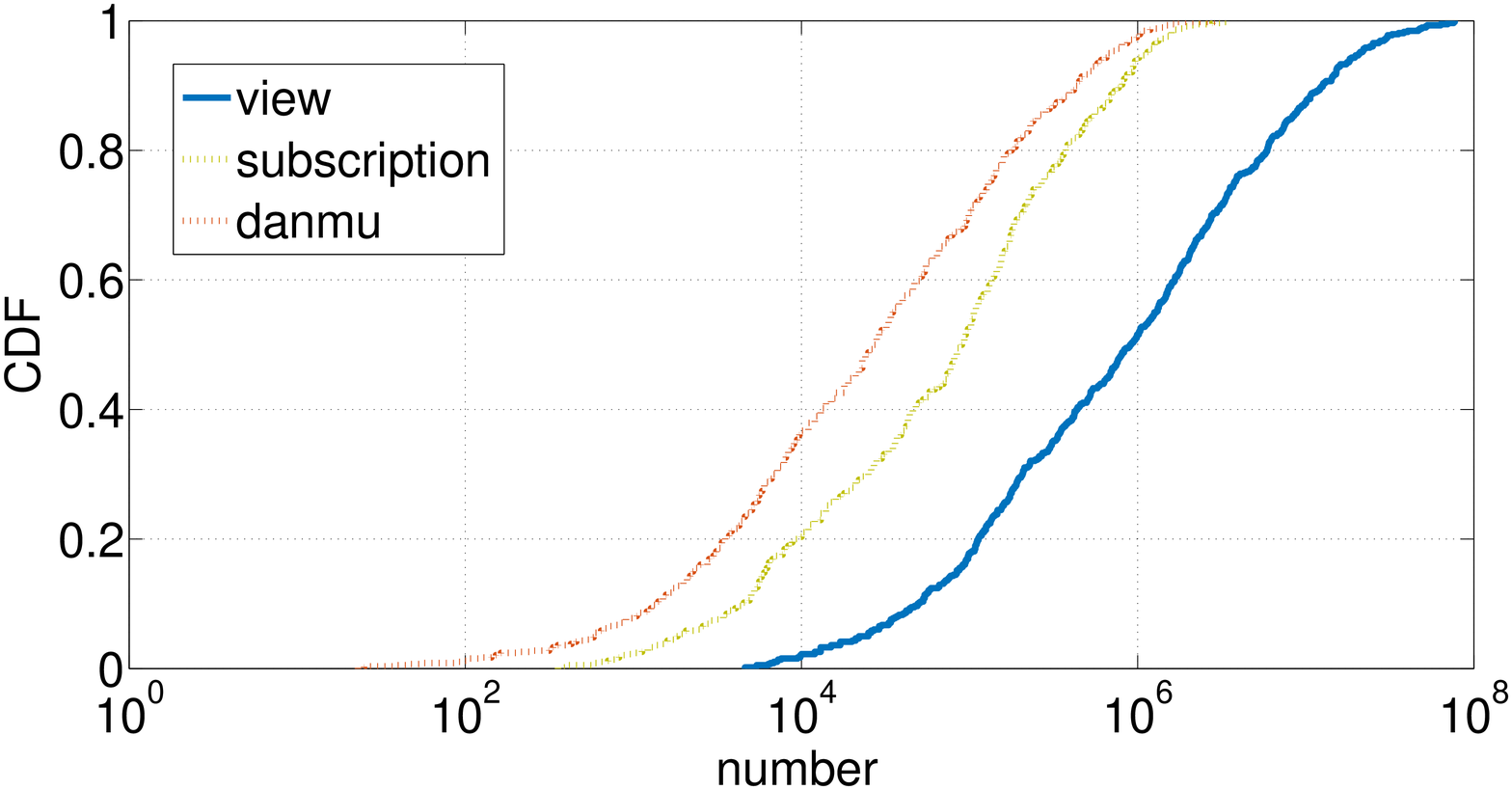} 
 \caption{CDFs of the number of views, the number of subscriptions, and the number of danmus collected by each copyright video series.}
  \label{cdf_juji}
\end{figure}

In this section, we reveal, in both the video level and the user level, the basic characteristics of user donations in BiliBili. The analysis will provide insights into the donation prediction and recommendation problem analyzed later in Section \ref{prediction}.

\subsection{The copyright video series}

We first examine the basic characteristics of all the 2,678 copyright video series purchased and published by the BiliBili community administrator by the end of 2017.

\textbf{Popyularity.} Figure \ref{cdf_juji} shows the cumulative distribution function (CDF) of the popularity of the copyright video series in terms of the number of views, the number of subscriptions, and the number of danmus they collect. Consistent with previous analyses \cite{YouTubeMoon, YouTubeUploader, Twitch12, GameReplayAdele}, it is more common for users to view than to subscribe and to leave danmus to the videos. On average, the number of views collected by the videos are one order of magnitude larger than the number of subscriptions and the number of danmus they collected. More specifically, 50\% video series have achieved over 1 million views whereas 60\% video series have attracted fewer than a hundred thousand subscriptions and 20 thousand danmus. These results indicate that as in many other video systems like YouTube and Twitch, many users in BiliBili are silent viewers.

Further, we find that for any metric, the popularity is highly skewed and a small fraction of the copyright video series achieve popularities that are orders of magnitude larger than the rest. Taking the number of views for example, the bottom 10\% video series have collected fewer than 20 thousand views whereas the top 20\% video series have collected over 10 million views. The highly skewed video popularity has been revealed in other video systems as well \cite{YouTubeMoon, YouTubeUploader, Twitch12, GameReplayAdele}.

\begin{figure}
 \centering
    \includegraphics[width=0.4\textwidth]{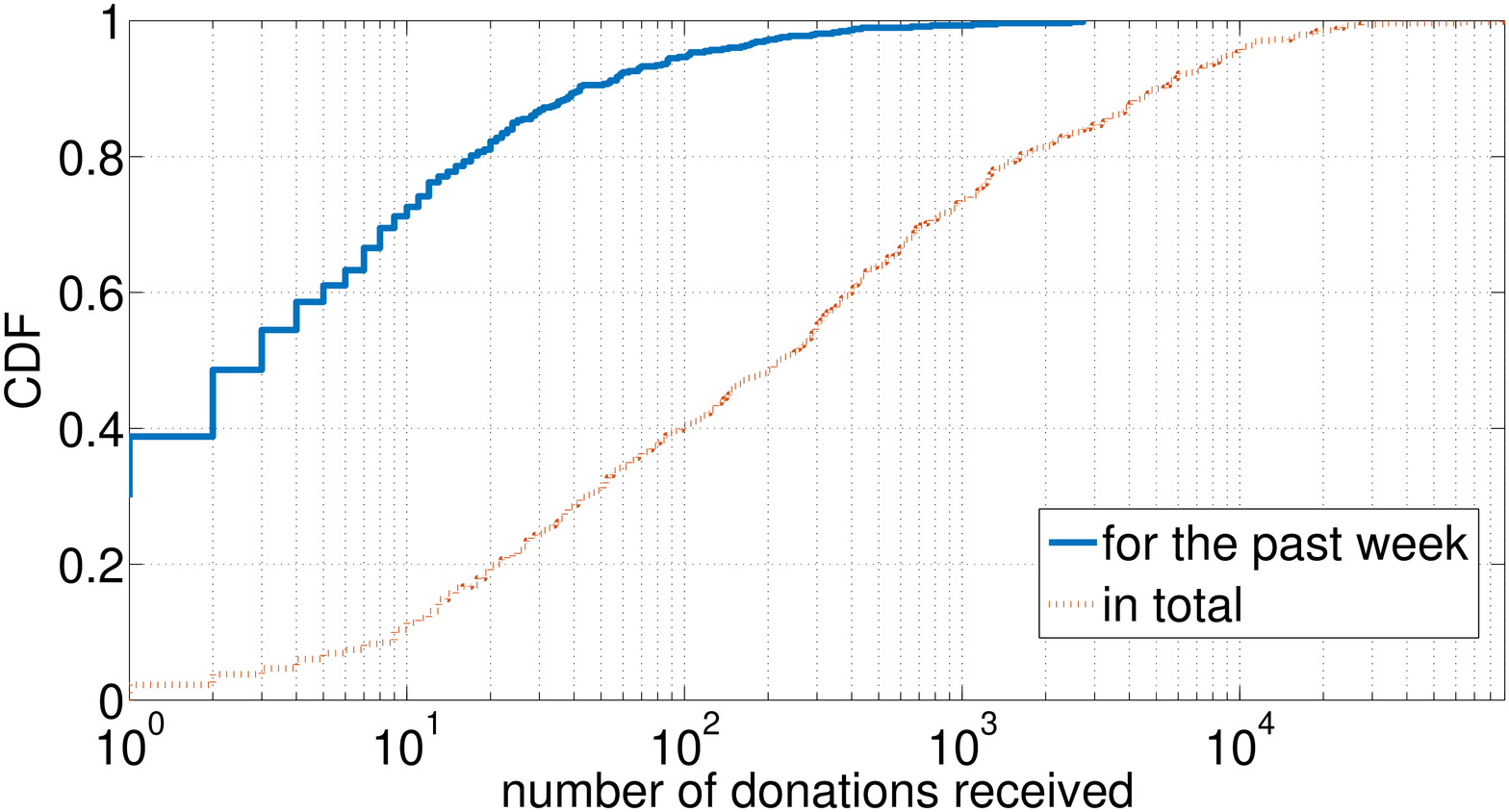}
      \caption{CDFs of the number of donations received by each copyright video series, for the past week and in total.}
  \label{cdf_sumCharge_juji}
\end{figure}

\textbf{Donation.} Similarly, as shown in Figure \ref{cdf_sumCharge_juji}, we also observe an uneven distribution for the number of donations received by each video series, for both the total donations and the donations received within one week before our crawling.
Specifically, while 50\% copyright video series have received fewer than 200 donations, around 30\% video series have received over a thousand donations in total. 3\% of them even managed to attract more than 200 donations in one week. 

We also notice that 40\% video series did not receive any donations for the past week.
One might argue that the number of donations received highly depends on the time when the video series are made available and that the 40\% video series with no new donations for the past week could be the very old ones.
Nevertheless, this is not the case. We have calculated the Spearman Ranking Correlation Coefficient (SRCC)\footnote{In brief, SRCC assesses how well the relationship between two variables can be described using a monotonic function \cite{Spearman}} between the video age and the number of donations they have received. For the total donation and donations received for the past week, they achieve a SRCC of 0.1507 and 0.0487, respectively. This result indicates that, counterintuitively, the number of donations are not correlated with the age of the video.

We further examine the correlations and calculated the SRCCs between the popularity metrics and the number of donations received.
Results are shown in Table \ref{srcc_juji}. We find that (i) the number of donations received in total is reasonably correlated with the three popularity metrics, each achieving a SRCC of over 0.75; and (ii) there are no strong correlations between the number of donations received for the past week and the popularity metrics, indicating that it will be difficult to infer or to predict future donations based solely on a single metric of the past popularity. Later in Section \ref{prediction}, we build machine-learned classifiers to solve this problem.



\begin{table}
\small
\centering
\caption{Spearman Ranking Correlation Coefficients (SRCC) between the age, the popularities and the number of donations received by the copyright video series.}
\label{srcc_juji}
\begin{tabular}{r|r|r|r|r}
\hline
 \textbf{SRCC}          & \textbf{$\#$ views} 	& \textbf{$\#$ subscriptions}	& \textbf{$\#$ danmus}  & \textbf{age} \\ \hline
	                
$\#$ total donations                       & 0.7876 & 0.8318 & 0.7688 & 0.1507 \\ 
$\#$ donations last week           & 0.5830 & 0.6925 & 0.6157 & 0.0487\\ 
 \hline
\end{tabular}
\end{table}

\subsection{Donators}

\begin{figure}
    \centering
    \includegraphics[width=0.4\textwidth]{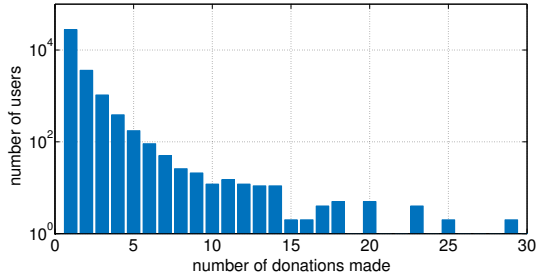}
      \caption{Number of donations to the copyright video series made by each user.}
  \label{bar_chargedVideo}
\end{figure}

In this section, we examine user activities of the 34,409 users who have donated at least once to the copyright video series.

\textbf{Donation.} Figure \ref{bar_chargedVideo} shows the distribution of the number of donations made by each donators. Consistent with our intuition, most users only donate to one copyright video series. 
We conjecture that they are the fans and that through donations they show their gratitude towards BiliBili for providing the copyright video series they like.
On the other hand, we find 85 users have donated to more than 10 copyright video series. 
They could be heavy consumers, i.e., being interested in a large number of video series, and through donations they encourage the community to publish more. It is also possible that these users are simply showing their support to the community. 
Reasoning on this behavior requires qualitative analyses such as interviews and surveys, which we leave as our future work.

\begin{figure}
 \centering
    \includegraphics[width=0.4\textwidth]{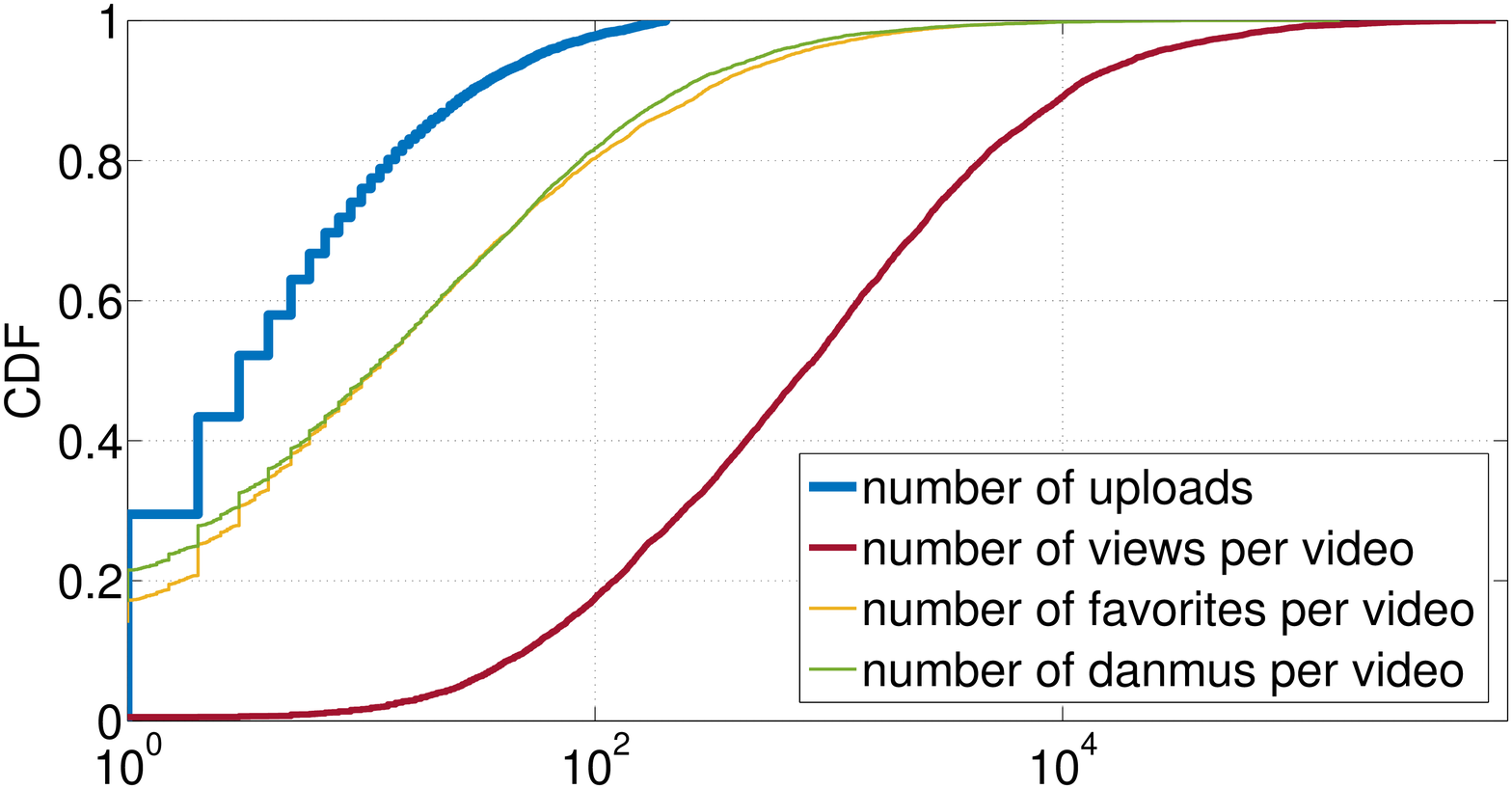}
      \caption{CDFs of the number of videos uploaded by donators, and the number of views, favorites, danmus, and comments they collected.}
  \label{cdf_chengbaoren_uploader}
\end{figure}

\textbf{Upload.} In total, 6,876 out of 34,409 (20\%) donators have not only donated money but also contributed UGC videos. We name them \textit{donators with uploads}.
Figure \ref{cdf_chengbaoren_uploader} shows the CDFs of the number of videos they have uploaded, and the average popularities of their videos in terms of the number of views, the number of favorites, and the number of danmus averaged over all the videos uploaded by each of them.
We find that 30\% of the donators with uploads have uploaded only once, whereas 20\% of them have uploaded more than 10 videos.
And the average popularities of their videos are also skewed: 10\% (20\%) of the donators with uploads collect on average over ten thousand views (fewer than 10 views) for each video they upload.

\begin{figure}
 \centering
    \includegraphics[width=0.4\textwidth]{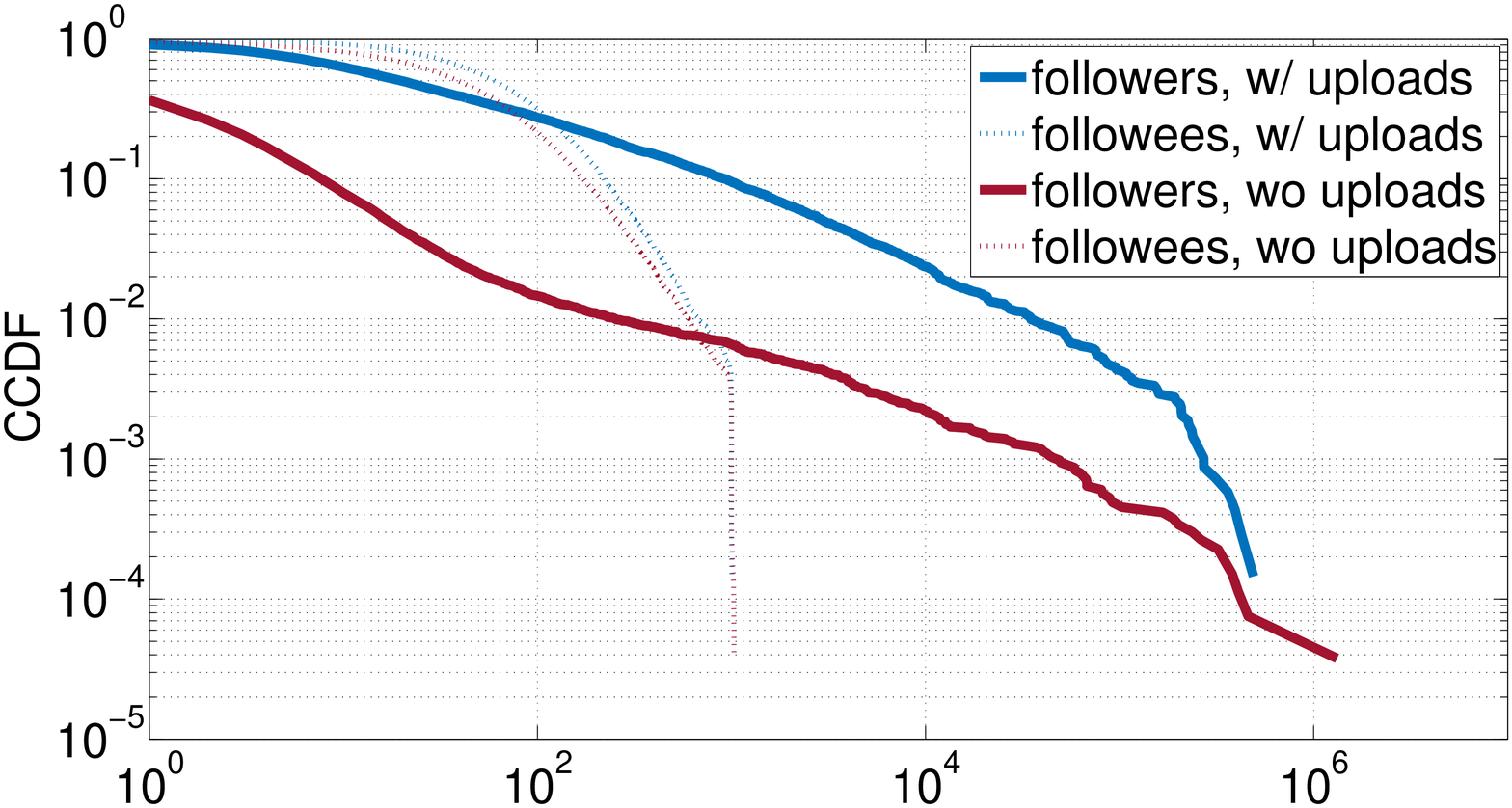} 
 \caption{CCDFs of the number of followers and the number of followees of donators.}
  \label{ccdf_chengbaoren_follow}
\end{figure}

\textbf{Follow.} 97\% of the donators have followed or have been followed by at least another user. In Figure \ref{ccdf_chengbaoren_follow} we show the complementary cumulative distribution function (CCDF) of the number of followers and the number of followees of the donators. 
Here, we have differentiated donators with (the blue lines) and without (the red lines) uploads, since intuitively users are more inclined to follow other users who have shared some videos. 

Consistent with our intuition, donators with uploads attract a lot more followers than donators with no uploads.
In total, nearly half of the donators with no uploads have never attracted any followers and over 80\% of them have fewer than 5 followers\footnote{We also find 386 of them have more than 100 followers. We have manually checked 10 of them and we found that they used to upload some videos which were later deleted due to various reasons.}, whereas on average donators with uploads attract 1,930 followers.
In the case of followees, donators with and without uploads achieve similar patterns, with a slightly higher number for the former one than the latter one.

Further, for both donators with and without uploads, the distributions of the number of followers exhibit straight lines when plotted on a log-log scale, indicating that they can be fitted properly with power-law distributions. On the other hand, the distributions of the number of followees exhibit a sudden drop on the tails, which is due to the limit imposed by the community on the number of users one can follow.

\begin{figure}
 \centering
    \includegraphics[width=0.4\textwidth]{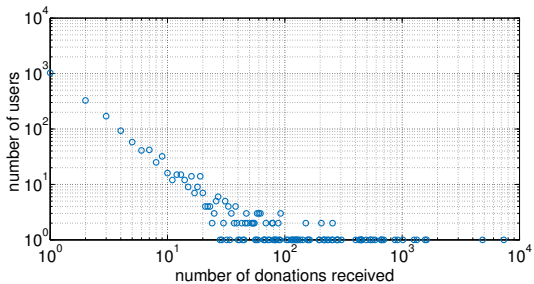}
      \caption{Number of donations from other users received by donators}
  \label{cdf_chargee_sumCharge}
\end{figure}

\begin{table*}[t]
\small
\centering
\caption{Performance evaluation of donation prediction}
\vspace{0.1cm}
\label{classification_results}
\begin{tabular}{l | l| l}
\hline
feature group  & description& AUC  \\ \hline

past popularity & the number of views, the number of danmus, and&  0.8021\\
 & the number of subscriptions received previously&  \\ \hline

past donation& the number of donations received previously and &  0.9762\\
& the number of donations received last week&  \\

 \hline
\end{tabular}
\end{table*}

\textbf{User-user donation.} Finally, a small fraction (3\%) of the donators have also received donations from other users for the UGC videos they share. We show in Figure \ref{cdf_chargee_sumCharge} the number of user-user donations received by the donators.
While most of them have only received one donation, some of them managed to attract hundreds and even thousands of donations for the UGC videos they upload.
Recall that in Figure \ref{cdf_sumCharge_juji} we have shown that 70\% copyright video series have received fewer than one thousand donations. 
These results suggest that users in BiliBili are willing to show their support not only for copyright contents, which they should pay anyway outside BiliBili, but also for UGC contents, which sometimes are appreciated even more than copyright videos.

In the above analysis, we have only considered user-user donations made to users who have donated to copyright video series. 
As our on-going work, the dataset and the analysis of user-user donations (with over 10 million videos and over 200 million users) will be publicly available soon.

\textbf{Discussion.} The above results reveal the prosperity of the BiliBili community, in video services, in social interactions, and in user donations. Our statistics indicate that in its own ways BiliBili has formed a very supportive community and are able to attract a considerable amount of user donations.
As a first step to argue the reasons for users to donate, we have calculated SRCCs between the number of donations made to copyright video series and the number of uploads (along with the average number of views, the average number of favorites, and the average number of danmus received), the number of followers, the number of followees, and the number of user-user donations received from other users. 
None of them has achieved a SRCC of over 0.1000, indicating that these metrics are not correlated and we cannot simply infer user donations from these metric alone. Later in Section \ref{prediction}, we apply network representation learning models to solve this problem.

\section{Donation prediction and recommendation}
\label{prediction}


Having gained several valuable insights on the characteristics of user donation in BiliBili, we now apply these findings and propose two models to mine the donation relationships, and we answer two fundamental questions in user donation, i.e., \textit{where the majority of donations go} and \textit{who will donate to whom}.
These questions will help the community to identify video series that will bring back large revenue and to find potential donators.

In the case of user donation to copyright video series in BiliBili, we form the two questions mentioned above as a prediction problem and a recommendation problem, respectively.

\textbf{Donation prediction}: with the current information obtained, can we predict the video series that will attract the majority of donations in the coming week?

\textbf{Donation recommendation}: for each copyright video series, can we recommend the potential users that will donate to them?




\subsection{Donation prediction}

\textbf{Experimental setup.} For our prediction task, we keep a record of all the copyright video series that have received at least one donation by the end of 2017 and we follow them for one week. We label the video series that receive top 20\% donations during the observation period as the positive examples and the rest as the negative examples. We train machine-learned classifiers based on two groups of features, i.e., \textit{past popularity} features including the number of views, the number of subscriptions, and the number of danmus received until the end of 2017, and \textit{past donation} features including the number of donations received in total and for the week before the end of 2017. 

We experimented with a variety of classification algorithms including svm, logistic regression, and random forests, and found the last one \cite{RandomForest} work best for our task. 
Hence all results reported here were obtained using random forests. 
For each experiment, we 
report the area under the receiver operating characteristic (ROC) curve (AUC).
We use balanced training and test sets containing equal numbers of positive and negative examples, so random guessing results in an AUC of 50\%. 

\textbf{Evaluation.} Results are shown in Table \ref{classification_results}. As it turns out, we can achieve a reasonable AUC of over 80\% using the past popularity features alone and together with the past donation features, the AUC reaches 97\%. The top three most informative features are the total number of donations received previously, the number of donations received last week, and the number of subscriptions received previously. It is interesting to notice that, the number of subscriptions is more informative for predicting donations received than the other two popularity metrics, i.e., the number of views and the number of danmus. We conjecture that viewing a video and leaving comments to it do not directly reflect user appreciation. For the former case, users may simply be exploring, or they may dislike it and will quickly turn it off. And for the latter case, it is possible that the video is a ``buzz'' maker and attracts a lot of attention and conflicts from factional users.

\begin{table*}[t]
\small
\centering
\caption{Performance evaluation of donation recommendation}
\label{rec_results}
\begin{tabular}{l|rr|rr|rr|rr}
\hline

 & MAP & recall@5 & MAP & recall@20  & MAP & recall@50&  MAP & recall@100 \\ \hline
 PMF & 0.3840 & 0.3884 & 0.4116 & 0.4197 & 0.4140 & 0.4279 & 0.4145 & 0.4498 \\
 CMF & 0.4109 & 0.4152 & 0.4461 & 0.4507 & 0.4501 & 0.4573 & 0.4507 & 0.4698 \\ \hline
metapath2vec &0.4176 & 0.4187 & 0.4452 & 0.4520 & 0.4513 & 0.4595  & 0.4515 & 0.4715  \\ 
HIN2vec &  0.4222 & 0.4244 & 0.4510 & 0.4645 & 0.4578 & 0.4870 & 0.4583 & 0.4999\\
\hline
\end{tabular}
\end{table*}

\subsection{Donation recommendation}

For our recommendation tasks, we first model BiliBili as a Heterogeneous Information Network (HIN) that consists of multiple types of nodes and edges and we adopt representation learning methods to learn the low-dimensional vector representations of the nodes and of the edges. Then, we use the learned representation vectors as input features and build machine-learned classifiers to rank and to recommend potential donators.

Recently, network representation learning (NRL) models are shown to perform well in link prediction and recommendation tasks \cite{NRLsurveyLeskovec}. They generate node sequences according to a variety of random walks and apply neural network models to learn the low-dimensional vector representations of the nodes to encode the structural information of the original network. For our experiment, we apply and compare two state-of-art NRL models, namely \textit{metapath2vec} \cite{metapath2vec} and \textit{HIN2vec} \cite{HIN2vec}. They are both specially designed for HINs.


\textbf{Experimental setup.} Similar to the above prediction task, we keep a record of all the copyright video series that have received at least one donation by the end of 2017 and their donators, and we follow them for one week.
We build a heterogeneous information network based on the \textit{user-video donation} and the \textit{user-user follow} relationships established by the end of 2017, which contains two types of nodes, i.e., user and video, and two types of edges, i.e., user-video donation (U-V) and user-user following (U-U). 
In total, this network contains 634 videos, 185,685 users (including 34,409 donators), 44,916 donation relationships, and 2,190,837 follow relationships.

We apply NRL models on this network to learn the vector representation for each user and for each video.
We have used the default parameters for the NRL models and ``U-U-V-U-U'' as the meta-path, i.e., the relation sequence connecting node pairs, to guide the random walks.
Then, the representation vector of a user-video pair is calculated by applying the Hadamard operator over the representation vectors of the corresponding two nodes (a user and a video). The Hadamard operator has been shown to perform stably and well for link prediction tasks using representation learning models \cite{node2vec}. In this way, we learn the vector representations for both the existing and the non-existing edges.

We label all the user-video donation relationships established during our observation period as the positive examples, and the user-video pairs with no edges as the negative examples.
Using the learned representations as input features,  
we apply the random forest algorithm to train a prediction model to rank node pairs that are more likely to have missing edges. To perform the recommendation, for each selected video we use its neighbours within 2 hops as the recommendation candidate.

\textbf{Evaluation.} Classic representation learning methods, especially for the recommendation problem, often rely on matrix factorization that factorizes the relationship matrix into low rank latent matrices and the nodes into low-dimensional vectors. Here, we compare the performance of NRL models with the following two matrix factorization methods.

\begin{itemize}
\item PMF: Probabilistic Matrix Factorization \cite{PMF} is the basic matrix factorization method using only user-item matrix. In our experiment, the matrix is constructed in a way that, when a user $i$ donates to a video $j$, the corresponding element $a_{ij}$ in the matrix is set to 1.
\item CMF: Collective Matrix Factorization  \cite{CMF} is a matrix factorization model that jointly factorizes different types of relations in HIN and shares the latent factor of same node types in different relations. In our experiment, we utilize a user-video matrix (constructed in the same way as in the above PMF experiment) and a user-user matrix that models user follow relationships (with an element $a_{ij}=1$ when a user $i$ and a user $j$ establish a follow relationship).
\end{itemize}

%

We use Mean Average Precision (MAP) and the top-k recall (recall@k) as metrics for evaluation, which are defined as follows respectively: 

\begin{itemize}
\item Mean Average Precision (MAP): the mean of the average precision score (AveP) for the ranking result of node $i$, where AveP($i$) = $\sum_{j=1}^k \frac{precision@j}{j}$ and MAP = $\frac{\sum_i^N AveP(i)}{N}$.
Here, precision@j represents the percentage of top-j ranking results that hit the ground truth and N represents the total number of nodes.
\item Top-k recall (recall@k): the fraction of the ground truth ranked in the top-k returned results.

recall@k = $\frac{\# hits\ in\ top-k}{\# positive\ examples\ in\ the\ ground\ truth}$

\end{itemize}

\vspace{+0.1cm}


\textbf{Results.} The results are shown in Table ~\ref{rec_results}.
It should be noted that PMF uses only user-video donation relationships and performs as the baseline for comparison. CMF and the two NRL models, on the other hand, use both user-video donation and user-user follow relationships. 
We have a number of interesting findings as follows:

Firstly, for any value of k, CMF and the two NRL models significantly outperform PMF (up to 10\% for MAP and 12\% for recall), indicating that integrating heterogeneous information improves the recommendation performance and that the additional social relationships are very informative for the donation recommendation problem.

Secondly, comparing NRL models with matrix factorization methods, the former always achieve better performance than the latter, even when they use the same amount of information. This result suggests that NRL models have better mechanism to integrate heterogeneous information.

Finally, between the two NRL models, HIN2vec performs better than metapath2vec. We believe that the additional improvement is achieved as HIN2vec learn latent vectors of nodes and meta-paths in the HIN simultaneously.


%

\section{Related work}

We summarize related work within each research topic our work covers as follows.

\textbf{Crowdsourced video systems.} Crowdsourced video systems like YouTube and Twitch.tv have been extensively studied before. 
Cha \textit{et al.} presented a comprehensive analysis of the popularity distribution and the time evolution of UGC video requests and their implications \cite{YouTubeMoon}.
Ding \textit{et al.} analyzed in-depth the behaviors of YouTube uploaders \cite{YouTubeUploader}.
Gill \textit{et al.} investigated YouTube from the perspective of YouTube traffic \cite{YouTubeTraffic}. They examined YouTube usage patterns, file properties, and transfer characteristics. 
Kaytoue \textit{et al.} provided preliminary characterizations on Twitch. They analyzed the dynamics of game spectators and proposed models for predicting video popularity \cite{Twitch12}. 
Jia \textit{et al.} compared Twitch with other systems, and investigated their repositories and user activities \cite{GameReplayAdele}. 
Wattenhofer \textit{et al.} analyzed the correlations between the popularity of YouTube videos and the properties of various social graphs created among the users \cite{YouTubeGraphs}.
Our analysis of video systems focuses on user donation, which to the best of our knowledge is still unexplored.

\textbf{Crowdfunding systems.}
On the other hand, user donation in crowdfunding platforms, wherein entrepreneurs solicit funding in order to bring their business plans, have been analyzed before, ranging from predicting the success of crowdfunding campaigns \cite{crowdfunding-chi13, crowdfunding-chi14, crowdfunding-osn, crowdfunding-15} to investor and project recommendation \cite{crowdfunding-rec-www14, crowdfunding-rec-aaai15, crowdfunding-rec-www15}, group recommendation \cite{crowdfunding-grouprec}, donator retention \cite{crowdfunding-donatorretention} and competition modelling \cite{crowdfunding-competition}. They mainly rely on probabilistic generative models or manual feature engineering (based on profile and social features) to build machine-learned classifiers to predict the success and the potential investors of the projects. 

In contrast, based on detailed donation relationships (who donated to whom) we build a heterogeneous information network and apply network representation learning models to automatically learn the node and the edge features.
Another major difference is that the above studies mainly focus on crowdfunding systems in the context of raising money for commercial projects, wherein donators are reimbursed by receiving interests or by pre-ordering the products. In BiliBili, however, donators do not expect any real return except for some videos and a friendly community. Thus, the donator dynamics and the donation relationships in BiliBili are expected to be different from the ones in crowdfunding systems and are worth investigating.

\textbf{Recommender systems.}
The donation recommendation problem analyzed in this article can be considered as a special case of recommender systems.
Recommender systems normally infer a user's preference (rating) for an item from historical interactions between users and items---the so-called collaborative filtering (CF) methods. Classic CF methods reply on matrix factorization to learn low-dimensional representations of the users and the items  \cite{PMF}, while additional information such as social relationships can be incorporated and jointly factorized \cite{CMF}. We refer the interested readers to \cite{recommenderSurveyTUD} for a comprehensive survey on recommender systems. Recently, as heterogeneous information networks (HINs) can naturally model complex relationships, a number of HIN-based methods are adopted in recommender systems. The basic idea is to improve user and item representations by utilizing meta-path based similarity, which can be calculated through matrix factorization \cite{hin-rec-wsdm14, SemRec, hin-rec-mf-kdd17} or random walks with learned edge weights \cite{hin-rec-randomwalks-wsdm18}.
 
A major difference between recommender systems and our recommendation task is that there is no explicit user ratings towards items (videos) in our case, and that instead of predicting the actual ratings, we focus on predicting user-video donation relationships. We believe, and also as shown in our experiments, that network representation learning models that originally designed for link prediction tasks are more suitable for our analysis.

\textbf{Network representation learning.} 
In recent years, inspired by the methods developed in image processing and natural language processing, there is a proliferation of network representation learning (NRL) models that learn low-dimensional representations of nodes based on neural network models. NRL models can effectively encode structural features and are shown to perform well in many data mining tasks including link prediction and node classification. 
A variety of NRL models have been proposed, for both homogeneous networks \cite{DeepWalk, LINE, node2vec} and heterogeneous networks \cite{HIN2vec, metapath2vec}. Extensive surveys on NRL models can be found here \cite{NRLsurvey-tkde18, NRLsurveyLeskovec}. 

It should be noted that in this work we do not focus on designing specific NRL models. Instead, for the first time to the best of our knowledge, we reveal the dynamics of user donations in crowdsourced video systems and the possibility of utilizing state-of-art representation learning methods to improve the donation recommendation problem.


\section{Conclusion and future work}

In this work, we conducted an analysis and presented the first publicly available dataset on user donation in crowdsourced video systems. Based on statistics on nearly 40 thousand users who have donated to copyright video series in Bilibili, we investigated the dynamics of user donations, and we applied our findings to accurately predict the destinations of the majority of the donations and for each video to recommend potential donators. 

We have a number of interesting findings. 
First, both the video popularity and the number of donations they received are highly skewed, with a small number of videos attracting a large number of views and donations.  And the user activity level is also highly skewed, in terms of the number of donations they made, the number of videos they upload, and the number of users who they follow and who follow them. 
Secondly and counterintuitively, we do not find any strong correlations between the video popularity and the number of donations received by the videos, nor between the number of donations made by the users and their upload and follow activity.
These results indicate that it is difficult to infer user donations solely from the manually engineered features.
Thirdly, when applying machine-learned classifiers to predict the future donations received by the videos, we find that the number of subscriptions received before is more informative than the number of views or comments, indicating that these popularity metrics reveal different type of popularities and that users indeed appreciate the video when they subscribe to it. 
Finally, for the donation recommendation problem, we find that social relationships provide complementary information and around 10\% performance improvements, and that when using the same amount of information, network representation learning models out-perform matrix factorization methods used in recommender systems.
Our on-going work on user-user donations in BiliBili covers a wider range of users, and the detailed information on their upload, social, and donation relationships. The dataset and the analysis (with over 10 million videos and over 200 million users) will be publicly available soon.


\bibliographystyle{ACM-Reference-Format}
\bibliography{socialcomputing}


\begin{thebibliography}{35}


\ifx \showCODEN    \undefined \def \showCODEN     #1{\unskip}     \fi
\ifx \showDOI      \undefined \def \showDOI       #1{#1}\fi
\ifx \showISBNx    \undefined \def \showISBNx     #1{\unskip}     \fi
\ifx \showISBNxiii \undefined \def \showISBNxiii  #1{\unskip}     \fi
\ifx \showISSN     \undefined \def \showISSN      #1{\unskip}     \fi
\ifx \showLCCN     \undefined \def \showLCCN      #1{\unskip}     \fi
\ifx \shownote     \undefined \def \shownote      #1{#1}          \fi
\ifx \showarticletitle \undefined \def \showarticletitle #1{#1}   \fi
\ifx \showURL      \undefined \def \showURL       {\relax}        \fi
\providecommand\bibfield[2]{#2}
\providecommand\bibinfo[2]{#2}
\providecommand\natexlab[1]{#1}
\providecommand\showeprint[2][]{arXiv:#2}

\bibitem[\protect\citeauthoryear{A.~Xu and Bailey}{A.~Xu and Bailey}{2014}]%
        {crowdfunding-chi14}
\bibfield{author}{\bibinfo{person}{H.~Rao W.-T. Fu S.-W.~Huang A.~Xu, X.~Yang}
  {and} \bibinfo{person}{B.~P. Bailey}.} \bibinfo{year}{2014}\natexlab{}.
\newblock \showarticletitle{Show me the money!: an analysis of project updates
  during crowdfunding campaigns}. In \bibinfo{booktitle}{\emph{{Proceedings of
  the SIGCHI conference on human factors in computing systems (CHI'14)}}}.
\newblock


\bibitem[\protect\citeauthoryear{Althoff and Leskovec}{Althoff and
  Leskovec}{2015}]%
        {crowdfunding-donatorretention}
\bibfield{author}{\bibinfo{person}{T. Althoff} {and} \bibinfo{person}{J.
  Leskovec}.} \bibinfo{year}{2015}\natexlab{}.
\newblock \showarticletitle{Donor Retention in Online Crowdfunding Communities:
  A Case Study of DonorsChoose.org}. In \bibinfo{booktitle}{\emph{{Proceeding
  of the 24th International World Wide Web Conference (WWW'15)}}}.
\newblock


\bibitem[\protect\citeauthoryear{Arantes, Figueiredo, and Almeida}{Arantes
  et~al\mbox{.}}{2016}]%
        {YouYubeAdsWebSci16}
\bibfield{author}{\bibinfo{person}{M. Arantes}, \bibinfo{person}{F.
  Figueiredo}, {and} \bibinfo{person}{J.M. Almeida}.}
  \bibinfo{year}{2016}\natexlab{}.
\newblock \showarticletitle{Understanding video-ad consumption on YouTube: a
  measurement study on user behavior, popularity, and content properties}. In
  \bibinfo{booktitle}{\emph{{Proceedings of the 8th ACM Conference on Web
  Science (WebSci'16)}}}.
\newblock


\bibitem[\protect\citeauthoryear{BiliBili}{BiliBili}{2017}]%
        {BiliBiliURL}
\bibfield{author}{\bibinfo{person}{BiliBili}.} \bibinfo{year}{2017}\natexlab{}.
\newblock
\newblock
\newblock
\shownote{https://www.bilibili.com.}


\bibitem[\protect\citeauthoryear{Breiman}{Breiman}{2001}]%
        {RandomForest}
\bibfield{author}{\bibinfo{person}{Leo Breiman}.}
  \bibinfo{year}{2001}\natexlab{}.
\newblock \showarticletitle{Random forests}.
\newblock \bibinfo{journal}{\emph{Machine Learning}}  \bibinfo{volume}{1}
  (\bibinfo{year}{2001}), \bibinfo{pages}{5--32}.
\newblock
Issue 45.


\bibitem[\protect\citeauthoryear{Cai, Zheng, and Chang}{Cai
  et~al\mbox{.}}{2018}]%
        {NRLsurvey-tkde18}
\bibfield{author}{\bibinfo{person}{H. Cai}, \bibinfo{person}{V.W. Zheng}, {and}
  \bibinfo{person}{K.C. Chang}.} \bibinfo{year}{2018}\natexlab{}.
\newblock \showarticletitle{A Comprehensive Survey of Graph Embedding:
  Problems, Techniques, and Applications}.
\newblock \bibinfo{journal}{\emph{IEEE transactions on knowledge and data
  engineering}} \bibinfo{volume}{30}, \bibinfo{number}{9}
  (\bibinfo{year}{2018}), \bibinfo{pages}{1616--1637}.
\newblock


\bibitem[\protect\citeauthoryear{Cha, Kwak, Rodriguez, Ahn, and Moon}{Cha
  et~al\mbox{.}}{2007}]%
        {YouTubeMoon}
\bibfield{author}{\bibinfo{person}{M. Cha}, \bibinfo{person}{H. Kwak},
  \bibinfo{person}{P. Rodriguez}, \bibinfo{person}{Y. Ahn}, {and}
  \bibinfo{person}{S. Moon}.} \bibinfo{year}{2007}\natexlab{}.
\newblock \showarticletitle{I Tube, You Tube, Everybody Tubes: Analyzing the
  World's Largest User Generated Content Video System}. In
  \bibinfo{booktitle}{\emph{{Internet measurement conference (IMC'07)}}}.
\newblock


\bibitem[\protect\citeauthoryear{Chung and Lee}{Chung and Lee}{2015}]%
        {crowdfunding-15}
\bibfield{author}{\bibinfo{person}{J. Chung} {and} \bibinfo{person}{K. Lee}.}
  \bibinfo{year}{2015}\natexlab{}.
\newblock \showarticletitle{A long-term study of a crowdfunding platform:
  Predicting project success and fundraising amount}. In
  \bibinfo{booktitle}{\emph{{Proceedings of the 26th ACM Conference on
  Hypertext $\&$ Social Media}}}.
\newblock


\bibitem[\protect\citeauthoryear{C.T.~Lu and Yu}{C.T.~Lu and Yu}{2014}]%
        {crowdfunding-osn}
\bibfield{author}{\bibinfo{person}{X.~Kong C.T.~Lu, S.~Xie} {and}
  \bibinfo{person}{P.~S. Yu}.} \bibinfo{year}{2014}\natexlab{}.
\newblock \showarticletitle{Inferring the impacts of social media on
  crowdfunding}. In \bibinfo{booktitle}{\emph{{Proceedings of the 7th ACM
  international conference on Web Search and Data Mining (WSDM'14)}}}.
\newblock


\bibitem[\protect\citeauthoryear{Ding, Du, Hu, Liu, Wang, Ross, and Ghose}{Ding
  et~al\mbox{.}}{2011}]%
        {YouTubeUploader}
\bibfield{author}{\bibinfo{person}{Y. Ding}, \bibinfo{person}{Y. Du},
  \bibinfo{person}{Y. Hu}, \bibinfo{person}{Z. Liu}, \bibinfo{person}{L. Wang},
  \bibinfo{person}{K.W. Ross}, {and} \bibinfo{person}{A. Ghose}.}
  \bibinfo{year}{2011}\natexlab{}.
\newblock \showarticletitle{Broadcast Yourself: Understanding YouTube
  Uploaders}. In \bibinfo{booktitle}{\emph{Proceedings of the 5th Internet
  Measurement Conference (IMC'11)}}.
\newblock


\bibitem[\protect\citeauthoryear{Dong, Chawla, and Swami}{Dong
  et~al\mbox{.}}{2017}]%
        {metapath2vec}
\bibfield{author}{\bibinfo{person}{Y. Dong}, \bibinfo{person}{N.V. Chawla},
  {and} \bibinfo{person}{A. Swami}.} \bibinfo{year}{2017}\natexlab{}.
\newblock \showarticletitle{metapath2vec: Scalable Representation Learning for
  Heterogeneous Networks}. In \bibinfo{booktitle}{\emph{{Proceeding of the 23th
  ACM SIGKDD international conference on Knowledge discovery and data mining
  (KDD'17)}}}.
\newblock


\bibitem[\protect\citeauthoryear{Fu, Lee, and Lei}{Fu et~al\mbox{.}}{2017}]%
        {HIN2vec}
\bibfield{author}{\bibinfo{person}{T.Y. Fu}, \bibinfo{person}{W.C. Lee}, {and}
  \bibinfo{person}{Z. Lei}.} \bibinfo{year}{2017}\natexlab{}.
\newblock \showarticletitle{{HIN2Vec: Explore Meta-paths in Heterogeneous
  Information Networks for Representation Learning}}. In
  \bibinfo{booktitle}{\emph{{Proceeding of the 26th international conference on
  Information and knowledge management (CIKM'17)}}}.
\newblock


\bibitem[\protect\citeauthoryear{Gill, Arlitt, Li, and Manhanti}{Gill
  et~al\mbox{.}}{2007}]%
        {YouTubeTraffic}
\bibfield{author}{\bibinfo{person}{P. Gill}, \bibinfo{person}{M. Arlitt},
  \bibinfo{person}{Z. Li}, {and} \bibinfo{person}{A. Manhanti}.}
  \bibinfo{year}{2007}\natexlab{}.
\newblock \showarticletitle{YouTube Traffic Characterization: A View from the
  Edge}. In \bibinfo{booktitle}{\emph{Proceedings of the first Internet
  Measurement Conference (IMC'07)}}.
\newblock


\bibitem[\protect\citeauthoryear{Greenberg, Hariharan, Gerber, and
  Pardo}{Greenberg et~al\mbox{.}}{2013}]%
        {crowdfunding-chi13}
\bibfield{author}{\bibinfo{person}{M. Greenberg}, \bibinfo{person}{K.
  Hariharan}, \bibinfo{person}{E. Gerber}, {and} \bibinfo{person}{B. Pardo}.}
  \bibinfo{year}{2013}\natexlab{}.
\newblock \showarticletitle{Crowdfunding Support Tools: Predicting Success \&
  Failure}. In \bibinfo{booktitle}{\emph{{Proceedings of the SIGCHI conference
  on human factors in computing systems (CHI'13)}}}.
\newblock


\bibitem[\protect\citeauthoryear{Grover and Leskovec}{Grover and
  Leskovec}{2016}]%
        {node2vec}
\bibfield{author}{\bibinfo{person}{A. Grover} {and} \bibinfo{person}{J.
  Leskovec}.} \bibinfo{year}{2016}\natexlab{}.
\newblock \showarticletitle{Node2vec: Scalable Feature Learning for Networks}.
  In \bibinfo{booktitle}{\emph{{Proceeding of the 22th ACM SIGKDD international
  conference on Knowledge discovery and data mining (KDD'16)}}}.
\newblock


\bibitem[\protect\citeauthoryear{Hamilton, Ying, and Leskovec}{Hamilton
  et~al\mbox{.}}{2018}]%
        {NRLsurveyLeskovec}
\bibfield{author}{\bibinfo{person}{W.L. Hamilton}, \bibinfo{person}{R. Ying},
  {and} \bibinfo{person}{J. Leskovec}.} \bibinfo{year}{2018}\natexlab{}.
\newblock \showarticletitle{Representation Learning on Graphs: Methods and
  Applications}. In \bibinfo{booktitle}{\emph{{arXiv:1709.05584}}}.
\newblock


\bibitem[\protect\citeauthoryear{J.~An and Crowcroft}{J.~An and
  Crowcroft}{2014}]%
        {crowdfunding-rec-www14}
\bibfield{author}{\bibinfo{person}{D.~Quercia J.~An} {and} \bibinfo{person}{J.
  Crowcroft}.} \bibinfo{year}{2014}\natexlab{}.
\newblock \showarticletitle{Recommending investors for crowdfunding projects}.
  In \bibinfo{booktitle}{\emph{{Proceeding of the 23th International World Wide
  Web Conference (WWW'14)}}}.
\newblock


\bibitem[\protect\citeauthoryear{Jia, Shen, Epema, and Iosup}{Jia
  et~al\mbox{.}}{2016}]%
        {GameReplayAdele}
\bibfield{author}{\bibinfo{person}{Adele~Lu Jia}, \bibinfo{person}{Siqi Shen},
  \bibinfo{person}{Dick Epema}, {and} \bibinfo{person}{Alexandru Iosup}.}
  \bibinfo{year}{2016}\natexlab{}.
\newblock \showarticletitle{When Game Becomes Life: The Creators and Spectators
  of Online Game Replays and Live Streaming}.
\newblock \bibinfo{journal}{\emph{ACM Trans. on Multimedia Computing,
  Communications, and Applications}} \bibinfo{volume}{12}, \bibinfo{number}{4}
  (\bibinfo{date}{August} \bibinfo{year}{2016}).
\newblock


\bibitem[\protect\citeauthoryear{Jiang, Liu, Fu, Wu, and Zhang}{Jiang
  et~al\mbox{.}}{2018}]%
        {hin-rec-randomwalks-wsdm18}
\bibfield{author}{\bibinfo{person}{Z. Jiang}, \bibinfo{person}{H. Liu},
  \bibinfo{person}{B. Fu}, \bibinfo{person}{Z. Wu}, {and} \bibinfo{person}{T.
  Zhang}.} \bibinfo{year}{2018}\natexlab{}.
\newblock \showarticletitle{Recommendation in Heterogeneous Information
  Networks Based on Generalized Random Walk Model and Bayesian Personalized
  Ranking}. In \bibinfo{booktitle}{\emph{{Proceedings of the 11th ACM
  international conference on Web search and data mining (WSDM'18)}}}.
\newblock


\bibitem[\protect\citeauthoryear{Kaytoue, Silva, and Raissi}{Kaytoue
  et~al\mbox{.}}{2012}]%
        {Twitch12}
\bibfield{author}{\bibinfo{person}{M. Kaytoue}, \bibinfo{person}{A. Silva},
  {and} \bibinfo{person}{C. Raissi}.} \bibinfo{year}{2012}\natexlab{}.
\newblock \showarticletitle{{Watch me Playing, I am a Professional: a First
  Study on Video Game Live Streaming}}. In
  \bibinfo{booktitle}{\emph{Proceedings of the 12th International World Wide
  Web Conference (WWW'12 Companion)}}.
\newblock


\bibitem[\protect\citeauthoryear{Krishnan and Sitaraman}{Krishnan and
  Sitaraman}{2013}]%
        {VideoAdsIMC13}
\bibfield{author}{\bibinfo{person}{S.~S. Krishnan} {and} \bibinfo{person}{R.~K.
  Sitaraman}.} \bibinfo{year}{2013}\natexlab{}.
\newblock \showarticletitle{Understanding the Effectiveness of Video Ads: A
  Measurement Study}. In \bibinfo{booktitle}{\emph{{Proceedings of the 13th
  Internet Measurement Conference (IMC'13)}}}.
\newblock


\bibitem[\protect\citeauthoryear{Lin, Yin, and Lee}{Lin et~al\mbox{.}}{2018}]%
        {crowdfunding-competition}
\bibfield{author}{\bibinfo{person}{Y. Lin}, \bibinfo{person}{P. Yin}, {and}
  \bibinfo{person}{W.C. Lee}.} \bibinfo{year}{2018}\natexlab{}.
\newblock \showarticletitle{Modeling Dynamic Competition on Crowdfunding
  Markets}. In \bibinfo{booktitle}{\emph{{Proceeding of the 27th International
  World Wide Web Conference (WWW'18)}}}.
\newblock


\bibitem[\protect\citeauthoryear{Perozzi, Al-Rfou, and Skiena.}{Perozzi
  et~al\mbox{.}}{2014}]%
        {DeepWalk}
\bibfield{author}{\bibinfo{person}{B. Perozzi}, \bibinfo{person}{R. Al-Rfou},
  {and} \bibinfo{person}{S. Skiena.}} \bibinfo{year}{2014}\natexlab{}.
\newblock \showarticletitle{DeepWalk: Online learning of social
  representations}. In \bibinfo{booktitle}{\emph{{Proceeding of the 20th ACM
  SIGKDD international conference on Knowledge discovery and data mining
  (KDD'14)}}}.
\newblock


\bibitem[\protect\citeauthoryear{Rakesh, Choo, and Reddy}{Rakesh
  et~al\mbox{.}}{2015a}]%
        {crowdfunding-rec-www15}
\bibfield{author}{\bibinfo{person}{V. Rakesh}, \bibinfo{person}{J. Choo}, {and}
  \bibinfo{person}{C.K. Reddy}.} \bibinfo{year}{2015}\natexlab{a}.
\newblock \showarticletitle{Project Recommendation Using Heterogeneous Traits
  in Crowdfunding}. In \bibinfo{booktitle}{\emph{{Proceeding of the 24th
  International World Wide Web Conference (WWW'15)}}}.
\newblock


\bibitem[\protect\citeauthoryear{Rakesh, Choo, and Reddy}{Rakesh
  et~al\mbox{.}}{2015b}]%
        {crowdfunding-rec-aaai15}
\bibfield{author}{\bibinfo{person}{V. Rakesh}, \bibinfo{person}{J. Choo}, {and}
  \bibinfo{person}{C.~K. Reddy}.} \bibinfo{year}{2015}\natexlab{b}.
\newblock \showarticletitle{Project recommendation using heterogeneous traits
  in crowdfunding}. In \bibinfo{booktitle}{\emph{{Proceedings of the 9th
  International AAAI Conference on Web and Social Media (CWSM'15)}}}.
\newblock


\bibitem[\protect\citeauthoryear{Salakhutdinov and Mnih}{Salakhutdinov and
  Mnih}{[n. d.]}]%
        {PMF}
\bibfield{author}{\bibinfo{person}{R. Salakhutdinov} {and} \bibinfo{person}{A.
  Mnih}.} \bibinfo{year}{[n. d.]}\natexlab{}.
\newblock \showarticletitle{Probabilistic matrix factorization}. In
  \bibinfo{booktitle}{\emph{Proceedings of the 20th International Conference on
  Neural Information Processing Systems (NIPS'08)}}.
\newblock


\bibitem[\protect\citeauthoryear{Shi, Zhang, Luo, Yu, Yue, and Wu}{Shi
  et~al\mbox{.}}{2015}]%
        {SemRec}
\bibfield{author}{\bibinfo{person}{C. Shi}, \bibinfo{person}{Z. Zhang},
  \bibinfo{person}{P. Luo}, \bibinfo{person}{P.S. Yu}, \bibinfo{person}{Y.
  Yue}, {and} \bibinfo{person}{B. Wu}.} \bibinfo{year}{2015}\natexlab{}.
\newblock \showarticletitle{{Semantic path based personalized recommendation on
  weighted heterogeneous information networks}}. In
  \bibinfo{booktitle}{\emph{{Proceeding of the 24th international conference on
  Information and knowledge management (CIKM'15)}}}.
\newblock


\bibitem[\protect\citeauthoryear{Shi, Larson, and Hanjalic}{Shi
  et~al\mbox{.}}{2014}]%
        {recommenderSurveyTUD}
\bibfield{author}{\bibinfo{person}{Y. Shi}, \bibinfo{person}{M. Larson}, {and}
  \bibinfo{person}{A. Hanjalic}.} \bibinfo{year}{2014}\natexlab{}.
\newblock \showarticletitle{Collaborative Filtering beyond the User-Item
  Matrix: A Survey of the State of the Art and Future Challenges}.
\newblock \bibinfo{journal}{\emph{Comput. Surveys}} (\bibinfo{year}{2014}).
\newblock


\bibitem[\protect\citeauthoryear{Singh and Gordon}{Singh and Gordon}{[n. d.]}]%
        {CMF}
\bibfield{author}{\bibinfo{person}{A.P. Singh} {and} \bibinfo{person}{G.J.
  Gordon}.} \bibinfo{year}{[n. d.]}\natexlab{}.
\newblock \showarticletitle{Relational learning via collective matrix
  factorization}. In \bibinfo{booktitle}{\emph{Proceeding of the 14th ACM
  SIGKDD international conference on Knowledge discovery and data mining
  (KDD'08)}}.
\newblock


\bibitem[\protect\citeauthoryear{Spearman}{Spearman}{1904}]%
        {Spearman}
\bibfield{author}{\bibinfo{person}{C. Spearman}.}
  \bibinfo{year}{1904}\natexlab{}.
\newblock \showarticletitle{The proof and measurement of association between
  two things}.
\newblock \bibinfo{journal}{\emph{American Journal of Psychology}}
  \bibinfo{volume}{15:72-101} (\bibinfo{year}{1904}), \bibinfo{pages}{72--101}.
\newblock


\bibitem[\protect\citeauthoryear{Tang, M.~Qu, Zhang, Yan, and Mei}{Tang
  et~al\mbox{.}}{2015}]%
        {LINE}
\bibfield{author}{\bibinfo{person}{J. Tang}, \bibinfo{person}{M.~Wang M.~Qu},
  \bibinfo{person}{M. Zhang}, \bibinfo{person}{J. Yan}, {and}
  \bibinfo{person}{Q. Mei}.} \bibinfo{year}{2015}\natexlab{}.
\newblock \showarticletitle{{LINE: Large-scale Information Network Embedding}}.
  In \bibinfo{booktitle}{\emph{Proceeding of the 22th International World Wide
  Web Conference (WWW'15)}}.
\newblock


\bibitem[\protect\citeauthoryear{V.~Rakesh and Reddy}{V.~Rakesh and
  Reddy}{2016}]%
        {crowdfunding-grouprec}
\bibfield{author}{\bibinfo{person}{W.-C.~Lee V.~Rakesh} {and}
  \bibinfo{person}{C.~K. Reddy}.} \bibinfo{year}{2016}\natexlab{}.
\newblock \showarticletitle{Probabilistic group recommendation model for
  crowdfunding domains}. In \bibinfo{booktitle}{\emph{{Proceedings of the 9th
  ACM International Conference on Web Search and Data Mining (WSDM'16)}}}.
\newblock


\bibitem[\protect\citeauthoryear{Wattenhofer, Wattenhofer, and Zhu}{Wattenhofer
  et~al\mbox{.}}{2012}]%
        {YouTubeGraphs}
\bibfield{author}{\bibinfo{person}{M. Wattenhofer}, \bibinfo{person}{R.
  Wattenhofer}, {and} \bibinfo{person}{Z. Zhu}.}
  \bibinfo{year}{2012}\natexlab{}.
\newblock \showarticletitle{The YouTube Social Network}. In
  \bibinfo{booktitle}{\emph{Proceedings of the 6th International AAAI
  Conference on Weblogs and Social Media (ICWSM'12)}}.
\newblock


\bibitem[\protect\citeauthoryear{Yu, Ren, Sun, Gu, Sturt, Khandel-wal, Norick,
  and Han.}{Yu et~al\mbox{.}}{2014}]%
        {hin-rec-wsdm14}
\bibfield{author}{\bibinfo{person}{X. Yu}, \bibinfo{person}{X. Ren},
  \bibinfo{person}{Y. Sun}, \bibinfo{person}{Q. Gu}, \bibinfo{person}{B.
  Sturt}, \bibinfo{person}{U. Khandel-wal}, \bibinfo{person}{B. Norick}, {and}
  \bibinfo{person}{Jiawei Han.}} \bibinfo{year}{2014}\natexlab{}.
\newblock \showarticletitle{Personalized entity recommendation: A heterogeneous
  information network approach}. In \bibinfo{booktitle}{\emph{{Proceedings of
  the 7th ACM international conference on Web search and data mining
  (WSDM'14)}}}.
\newblock


\bibitem[\protect\citeauthoryear{Zhao, Yao, Li, Song, and Lee}{Zhao
  et~al\mbox{.}}{2017}]%
        {hin-rec-mf-kdd17}
\bibfield{author}{\bibinfo{person}{H. Zhao}, \bibinfo{person}{Q. Yao},
  \bibinfo{person}{J. Li}, \bibinfo{person}{Y. Song}, {and} \bibinfo{person}{D.
  Lee}.} \bibinfo{year}{2017}\natexlab{}.
\newblock \showarticletitle{Meta-graph based recommendation fusion over
  heterogeneous information networks}. In \bibinfo{booktitle}{\emph{{Proceeding
  of the 23th ACM SIGKDD international conference on Knowledge discovery and
  data mining (KDD'17)}}}.
\newblock


\end{thebibliography}

\end{document}